\documentclass[mathleft
]{an}
\usepackage{graphicx}
\usepackage{times}
\def\note #1]{{\bf #1]}}
\def\titnt #1].{\hskip -0.1pt}
\def\dd{{\rm d}}
\def\bolddelta{\delta\kern-0.45em\delta\kern-0.45em\delta}

\def\CG{{\cal G}}
\def\CH{{\cal H}}
\def\CK{{\cal K}}
\def\CP{{\cal P}}

\def\nubest{\nu_{nl}^{\rm (best)}}
\def\nuref{\nu_{nl}^{\rm (ref)}}
\usepackage{natbib}
\bibpunct{(}{)}{;}{a}{}{,}
\begin{document}

\Pagespan{789}{}
\Yearpublication{2006}%
\Yearsubmission{2005}%
\Month{11}%
\Volume{999}%
\Issue{88}%


\title{Stellar model fits and inversions}

\author{J. Christensen-Dalsgaard\thanks{
  \email{jcd@phys.au.dk}\newline}
}
\titlerunning{Stellar model fits and inversions}
\authorrunning{J. Christensen-Dalsgaard}
\institute{
Stellar Astrophysics Centre, Department of Physics and Astronomy, 
Aarhus University, Ny Munkegade 120,\\
8000 Aarhus C, Denmark}

\received{??? September 2012}
\accepted{??? September 2012}
\publonline{later}


\abstract{%
The recent asteroseismic data from the CoRoT and {\it Kepler} missions
have provided an entirely new basis for investigating stellar properties.
This has led to a rapid development in techniques for analysing such data,
although it is probably fair to say that we are still far from having the tools
required for the full use of the potential of the observations.
Here I provide a brief overview of some of the issues related 
to the interpretation of asteroseismic data.
}

\maketitle

\section{Introduction}



The goals of the asteroseismic analysis are evidently to improve
our knowledge about stellar properties and our understanding of
stellar structure and evolution.
More specifically, simple model fits, or even direct use of scaling
relations, provide estimates of the masses and radii of stars, while
more detailed analyses, including also other observations of the stars,
yield more precise values of mass and radius as well as an estimate 
of the stellar age.
Such determinations clearly have substantial broad interest, not
least for stars known to be hosts of planetary systems.
However, from the point of view of stellar astrophysics we
clearly wish more information, and indeed we can go much further
from the data now becoming
available, not least from the CoRoT and {\it Kepler} missions
\citep{Michel2008, Baglin2009, Boruck2010, Gillil2010, Koch2010}.
Recent or upcoming highlights are the inferences on the core structure
and rotation of red-giant stars 
\citep[e.g.,][]{Beddin2011, Beck2012, Mosser2012a, Mosser2012b}
and the detection of effects in solar-like oscillations
of acoustic glitches in main-sequence stars 
\citep{Mazumd2012a, Mazumd2012b}.
Fortunately, both missions have now been extended until 2016.

Much of the data fitting or other form of analysis is done within
the framework of stellar models characterized by a specific set
of parameters.
A crucial goal of asteroseismology is to test whether such models
adequately represent real stars and, if not, how the model physics 
should be improved.
Inadequate modelling should be reflected in a failure of the models
to fit the observed stellar properties, particularly the frequencies,
to within their observational errors,
for any choice of parameters characterizing the model.
At this point one then has to determine how the models should be
further improved, perhaps through inverse analyses to try to
localize the cause of the discrepancy.
It is evident that the identification of such significant departures
requires great care in the statistical analysis of the data, and
it is probably fair to say that we have so far not quite reached this point.

In this brief review I concentrate on the determination of stellar
structure based on solar-like oscillations, with main emphasis on
stars on or near the main sequence.
However, it should be noted in passing that striking results have
also been obtained based on CoRoT and {\it Kepler} data for
more massive stars showing heat-engine driven pulsations
\citep{Degroo2010}
and for the subdwarf B stars \citep[e.g.,][]{VanGro2010}.
Also, extensive ground-based campaigns have yielded fundamental results
for white dwarfs \citep[e.g.,][]{Metcal2004}.
Thus asteroseismic investigations are covering a very broad range
of stellar properties and stellar evolutionary stages.

\section{Properties of stellar oscillations}


As a background for the discussion, I present a brief overview of
the relevant properties of stellar oscillations.
Much more detailed descriptions were provided, e.g., by
\citet{Christ2004} and \citet{Aerts2010}.

\subsection{Basic properties}

We consider only oscillations that are solar-like, in the sense
that they are intrinsically damped and excited by the acoustic 
noise generated by near-surface convection, which reaches near-sonic speeds
\citep[e.g.,][]{Goldre1977, Houdek1999}.
This mechanism depends on the near-surface properties of the 
oscillations and the time-dependence of the most vigorous convection
\citep{Kjelds2011},
making it most efficient at frequencies approaching the acoustic
cut-off frequency in the stellar atmosphere, approximately given,
in terms of cyclic frequency, by
\begin{equation}
\nu_{\rm ac} \simeq {c_{\rm phot} \over 4 \pi H_p} \propto
{G M \over R^2 T_{\rm eff}^{1/2}} \; ,
\end{equation}
where $c_{\rm phot}$ and $H_p$ are the sound speed and pressure scale height 
at the stellar photosphere.
The second expression assumes the ideal gas law and that
the atmospheric temperature scales like the effective temperature 
$T_{\rm eff}$;
here $G$ is the gravitational constant, and $M$ and $R$ are the mass
and radius of the star.

In relatively unevolved stars, on the main sequence, 
the modes in the frequency range of efficient driving are acoustic modes,
with cyclic frequencies that are asymptotically given by
\begin{equation}
\nu_{nl} \simeq \Delta \nu (n + l/2 + \epsilon) - \delta_{0l} 
\label{eq:pasymp}
\end{equation}
\citep{Vandak1967, Tassou1980, Gough1986},
where $n$ and $l$ are the radial order and spherical harmonic degree
of the mode.
Here $\Delta \nu$ is approximately the inverse acoustic diameter of the
star,
\begin{equation}
\Delta \nu = \left( 2 \int_0^R {\dd r \over c} \right)^{-1} \; ,
\end{equation}
where $c$ is the adiabatic sound speed and $r$ is the distance to the
centre of the star.
As expected from homology arguments this scales approximately
as the square root of the mean density of the star,
\begin{equation}
\Delta \nu \propto \left( {M \over R^3} \right)^{1/2} \; .
\label{eq:delnu}
\end{equation}
Also, $\epsilon$ is a phase depending on the near-surface properties
of the mode and $\delta_{0l}$ is a small correction that in
main-sequence stars is sensitive to the central regions of the star.

In terms of the observed frequencies the asymptotic properties
in Eq.~(\ref{eq:pasymp}) are often utilized by considering the
large frequency separation
\begin{equation}
\Delta \nu_{nl} = \nu_{nl} - \nu_{n-1 \, l} \; ,
\label{eq:largesep}
\end{equation}
and the small frequency separations
\begin{equation}
\delta \nu_{nl} = \nu_{nl} - \nu_{n-1 \, l+2} \; .
\label{eq:smallsep}
\end{equation}
Here $\Delta \nu_{nl}$ approximates $\Delta \nu$, 
although with some variations with frequency that largely arise from
the near-surface acoustic glitches (see Section \ref{sec:glitch});
$\delta \nu_{nl}$, as $\delta_{0l}$,
provides a measure of conditions in the core of the star and hence,
for main-sequence stars, the stellar age.

The amplitudes of the modes are determined by a balance between
the energy input from the stochastic excitation and the damping.
This leads to a bell-shaped distribution of amplitudes with
frequency \citep{Goldre1994} with a typically
relatively well-defined maximum at a frequency $\nu_{\rm max}$.
There is observational evidence that $\nu_{\rm max}$ scales
as $\nu_{\rm ac}$, i.e.,
\begin{equation}
\nu_{\rm max} \propto {G M \over R^2 T_{\rm eff}^{1/2}} 
\label{eq:numax}
\end{equation}
\citep[e.g.,][]{Brown1991, Beddin2003, Stello2008},
with some although perhaps not full theoretical justification in terms
of the near-surface properties of the modes \citep{Belkac2011}.

The large and small frequency separations and $\nu_{\rm max}$ form the
basis for ensemble studies of large samples of stars and hence are
typically determined by automated pipeline analysis methods.
A careful comparison of such techniques was provided by \citet{Verner2011a}.

As the star evolves beyond the main sequence, the core contracts strongly and
hence the internal gravitational acceleration becomes very large.
Furthermore, hydrogen burning leaves behind strong gradients in the chemical
composition.
These effects lead to a marked increase in the so-called buoyancy frequency
$N$,
which for an ideal gas can be approximated by
\begin{equation}
N^2 \simeq {g^2 \rho \over p} (\nabla_{\rm ad} - \nabla + \nabla_\mu) \; ,
\label{eq:buoy}
\end{equation}
where $p$ is pressure, $\rho$ density, 
$g$ is the local gravitational acceleration,
$\nabla = \dd \ln T / \dd \ln p$, $T$ being temperature, $\nabla_{\rm ad}$
is the adiabatic value of $\nabla$ and $\nabla_\mu = \dd \ln \mu / \dd \ln p$,
where $\mu$ is the mean molecular weight.
$N$ is the characteristic frequency for internal gravity waves, 
and when $N$ is large the star has {\it mixed modes}, which behave like
acoustic modes in the outer parts of the star and as gravity modes in the deep
interior. 
For sufficiently high $N$, found in sub-giant and red-giant stars, such mixed
modes have frequencies as high as, or exceeding, $\nu_{\rm ac}$ and hence 
are excited efficiently by the stochastic near-surface processes
\citep[e.g.,][]{Dupret2009}.
Given the dependence of $N$ on the structure of the deep interior of the star,
including the composition profile, these modes are powerful diagnostics
of the stellar interior.

High-order asymptotic g modes are uniformly spaced in period $\Pi_{nl}$, 
which asymptotically satisfies
\begin{equation}
\Pi_{nl} \simeq {\Pi_0 \over \sqrt{l(l+1)}} (n + \epsilon_{\rm g}) \; ,
\label{eq:gasymp}
\end{equation}
where
\begin{equation}
\Pi_0 = 2 \pi^2 \left( \int N {\dd r \over r} \right)^{-1} \; ,
\label{eq:perspac}
\end{equation}
the integral being over the region where the buoyancy frequency
exceeds the oscillation frequency, and $\epsilon_{\rm g}$ is a phase.
For red giants this leads to a very dense spectrum of high-order 
g modes interacting with the acoustic modes in the region of 
stochastic excitation.
For modes with $l = 1$ several such mixed modes are typically 
observed in the vicinity of each acoustic mode.
This allows a determination of the corresponding period spacing
$\Delta \Pi_1 = \Pi_0/\sqrt{2}$ which, according to Eq.~(\ref{eq:perspac}),
is a diagnostic of the deep stellar interior where $N$ is large
\citep{Beck2011, Beddin2011, Mosser2011}.

In a spherically symmetric star the frequencies do not depend on 
the azimuthal order $m$ of the spherical harmonic.
Rotation lifts this degeneracy, giving rise to a rotational splitting which
depends on the internal rotation of the star, weighted by the eigenfunction
of the mode.
Observations of the rotational splitting in the Sun have provided detailed 
information about the solar internal rotation
\citep[for a review, see][]{Thomps2003}, and {\it Kepler} has provided 
striking observations of the splitting of solar-like oscillations in 
evolved stars
\citep[e.g.,][]{Beck2012, Deheuv2012, Mosser2012b}.
However, here I ignore rotation and concentrate on the determination of 
stellar properties based on the average multiplet frequencies $\nu_{nl}$.

\subsection{Near-surface problems}
\label{sec:nearsurf}


Modelling of near-surface layers of a solar-like star is complicated by
the presence of convection.
The temperature structure of the model is determined by the treatment of
convective energy transport, often using the mixing-length formulation
\citep{Bohm1958},
and the models normally ignore the dynamical effects of convection, in
the so-called turbulent pressure.
Also, the frequency calculations are most often carried out in the adiabatic
approximation, even though the oscillations are strongly nonadiabatic in the 
superficial layers.
However, the treatment of the nonadiabatic effects needs a description of
the interaction between convection and pulsations, through the perturbation
to the convective flux; also, the perturbation to the turbulent pressure
may affect both the frequencies and the stability of the modes.
An overview of these complications was given by \citet{Houdek2010}.
It should be noticed that the uncertain aspects of the structure and
oscillation modelling are all concentrated very near the surface.
Thus they have little effect on modes of low frequency which are evanescent
in this region, with a very low amplitude compared with the amplitude in
the bulk of the star.

\begin{figure}
\includegraphics[width=80mm]{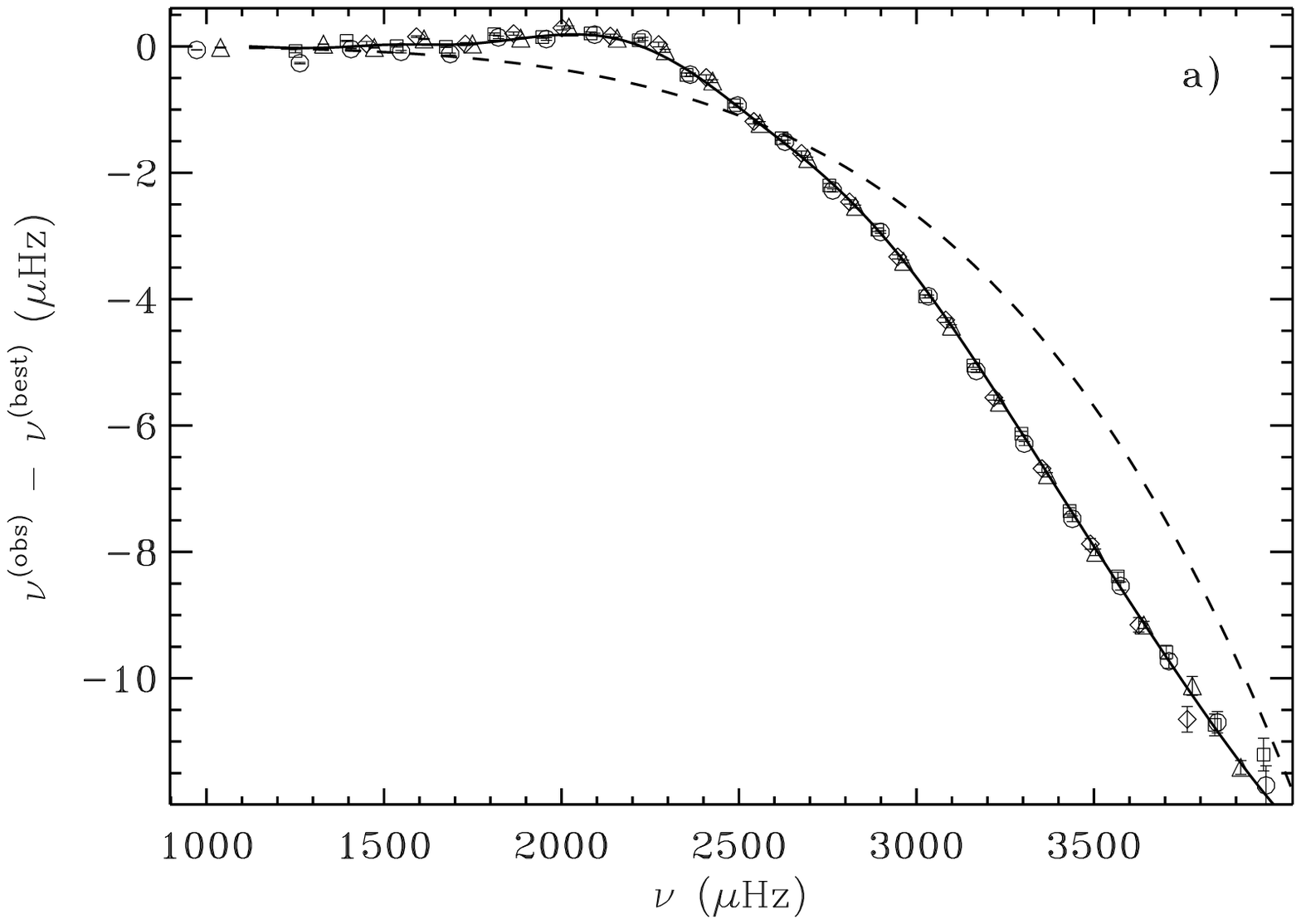}
\includegraphics[width=80mm]{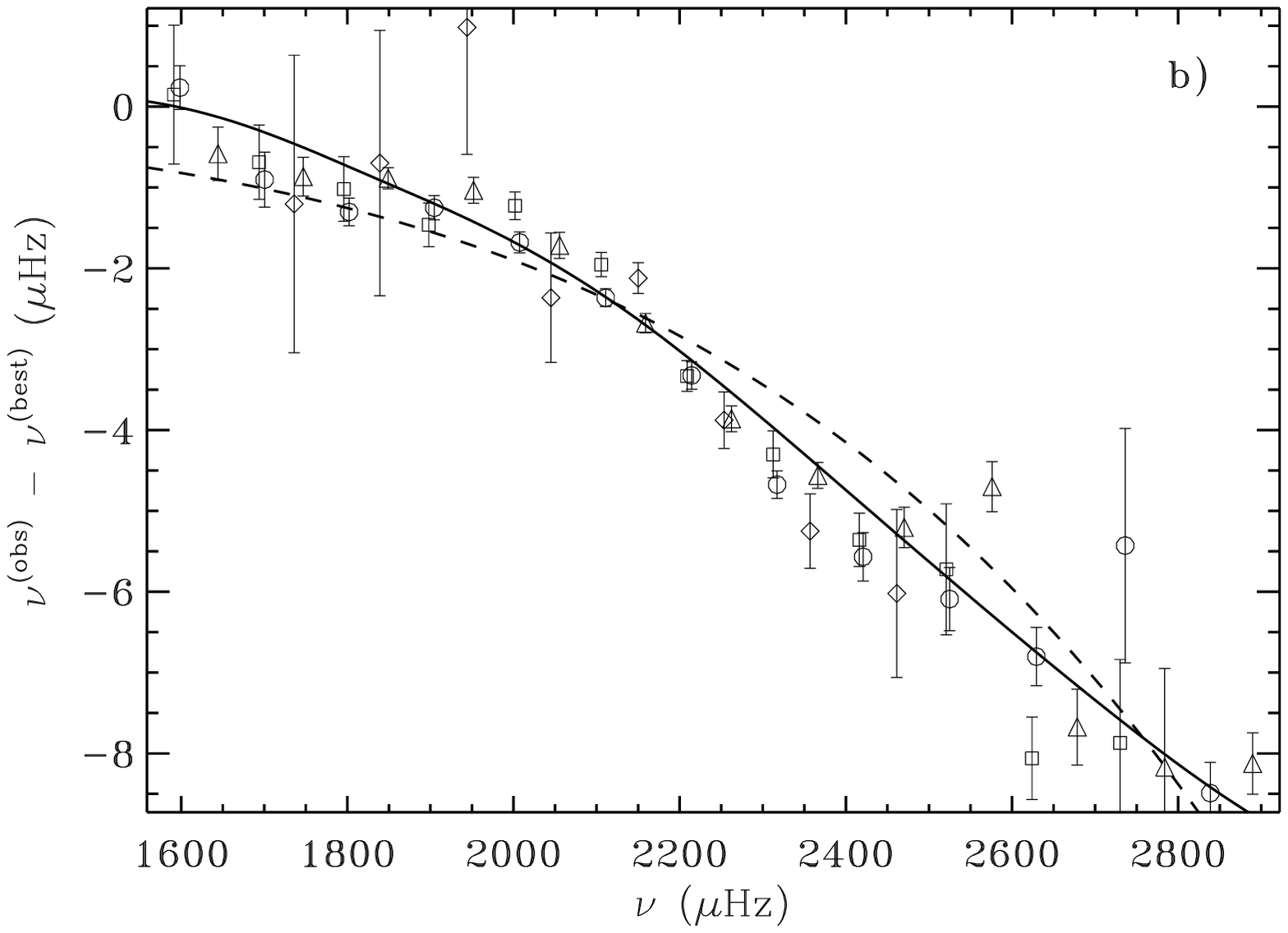}
\caption{
In panel a) the solid line shows the surface contribution to the
differences between the solar and model frequencies, inferred from
analysis of modes over a broad range in degree.
The symbols show the result of fitting this function to BiSON frequencies
\citep[see][]{Chapli2002} of degree $l = 0{-}3$. 
For comparison, the dashed line is the corresponding fit of a power law
(cf.\ Eq. \ref{eq:surface}) with exponent $b = 4.9$
In panel b) the symbols show the the differences between observed frequencies
of 16 Cyg A \citep{Metcal2012} and frequencies of a fitted model, without
the surface correction that was used in the fit.
That surface correction is shown by the dashed line, while the solid line
shows the scaled solar surface contribution from panel a), assumed to
be a function of the frequency in units of the acoustic cut-off frequency.
}
\label{fig:surf}
\end{figure}

These near-surface problems dominate the differences between observed solar
oscillation frequencies and frequencies of solar models \citep{Christ1996}.
In this case it is possible to separate the near-surface effects from
the differences between the Sun and the model in the deeper parts of
the star.
A convenient way to do so is to use the differential form of
the Duvall asymptotic expression for the frequencies
\citep{Christ1989}, according to which the frequency differences
$\delta \nu$ between the Sun and a model satisfy
\begin{equation}
S_{nl} {\delta \nu_{nl} \over \nu_{nl}} \simeq
\CH_1(\nu_{nl}/L) + \CH_2(\nu_{nl}) \; ,
\end{equation}
where $L = \sqrt{l(l+1)}$ and
$S_{nl}$ is a scale factor which can be calculated from the model,
and which may be chosen to be close to 1 for low-degree modes.
The term in $\CH_1$ depends on the location of the inner turning
point of the mode, determined by $\nu/L$, and reflects the difference
in sound speed between the Sun and the model throughout the star,
whereas $\CH_2$ contains the contribution from the near-surface region.
In accordance with the above comment the arbitrary additive constant 
in the definition of $\CH_2$ may be chosen such that $\CH_2$ is zero at
low frequency.
In Fig.~\ref{fig:surf} the solid curve shows $\nu \CH_2(\nu)$
determined in this manner, corresponding to the frequency
shift $\delta \nu$ caused by the surface effects in the Sun.

In principle, these effects must be taken into account in any analysis
of frequencies of solar-like oscillations.
They affect not only the individual frequencies but also frequency combinations
such as the large and small frequency separations
(cf.\ Eqs \ref{eq:largesep} and \ref{eq:smallsep}).
\citet{Roxbur2003a} pointed out that separation ratios, such as
\begin{equation} 
r_{02} = {\nu_{n0} - \nu_{n-1\,2} \over \nu_{n1} - \nu_{n-1\, 1}} \; ,
\end{equation}
can be constructed which are essentially independent of the near-surface
problems
\citep[see also][]{Oti2005, Roxbur2005}.
These combinations therefore provide clean diagnostics of the core properties
of the stars.

To fit individual frequencies it is necessary to correct for the 
near-surface effects.
For distant stars, where only low-degree modes are observed, 
an unconstrained determination of the frequency corrections cannot
be carried out.
\citet{Kjelds2008} suggested to base the correction on the solar results,
fitting the observed frequencies to
\begin{equation}
\nu_{nl} = \nubest + a \left( {\nu_{nl} \over \nu_0} \right)^b \; ,
\label{eq:surface}
\end{equation}
where $\nubest = r \nuref$ are the frequencies of the best-fitting model,
taking into account the near-surface correction,
and $\nuref$ are frequencies of a computed reference model, assumed to
be close to the best model.
The scaling with the constant factor $r$ assumes that the frequencies of
the models are related by homology.
In Eq.~(\ref{eq:surface}) the exponent $b$ is obtained from fits to solar 
frequencies, and $\nu_0$ is typically taken to be an average of the observed
frequencies, which is generally close to $\nu_{\rm max}$.
Kjeldsen et~al. provide a procedure for determining the amplitude $a$ 
of the correction and the frequency scale factor, by fitting the 
mean large frequency separation and average frequency for radial modes.

\citet{Mathur2012} carried out detailed fits to a number of stars observed
by {\it Kepler}, including the \citet{Kjelds2008} correction.
They found that the amplitudes $a$ roughly scaled as $\Delta \nu$.
According to Eq.~(\ref{eq:pasymp}) this corresponds to a roughly constant
average shift in the phase $\epsilon$.
A similar shift was found by \citet{White2011a} in a comparison of fitted
values of $\epsilon$ from observed and model frequencies.
This suggests that a more detailed analysis of the phase, including how it
is affected by the near-surface layers, may provide a deeper insight into
the properties of the frequency correction. 

Although the power-law correction assumed in Eq.~(\ref{eq:surface}) has
some theoretical support in an analysis of simple near-surface modifications
\citep{Christ1980} it is clearly only an approximation to the frequency
differences obtained for the Sun 
(see Fig.~\ref{fig:surf}).
An alternative is to base the correction more closely on the functional
form obtained in the solar case.
It is perhaps not unreasonable to assume that the relevant frequency
scale is the acoustic cut-off frequency $\nu_{\rm ac}$ and hence
to replace Eq.~(\ref{eq:surface}) by
\begin{equation}
\nu_{nl} = \nubest + \tilde a \CG_\odot(\nu_{nl}/\nu_{\rm ac}) \; ,
\label{eq:solsurf}
\end{equation}
where the function $\CG_\odot$ is determined from 
$\nu \CH_2(\nu)$, illustrated in Fig.~\ref{fig:surf},
as obtained from fitting solar frequencies over a broad range of
degrees.
The scale factor $\tilde a$ and the frequency scale $r$ can be determined
from a least-squares fit to the differences between the observed and the
reference-model frequencies.
In Fig.~\ref{fig:surf} the symbols illustrate the result of
applying this procedure to observed solar frequencies of degree
$l = 0{-}3$, corresponding to what can be observed in distant stars,
and using the same reference model as was used to determine $\CG_\odot$.
Here the scale factor was extremely close to 1.
For comparison, the dashed curve shows the result of applying the
\citet{Kjelds2008} procedure to the same set of frequencies, with the
same reference model.
It is evident that, while the power law reproduces the general rapid increase
of the effect with frequency, the detailed fit to the extensive and
very accurate solar frequencies is somewhat questionable.


As a more realistic illustration of the fit to the surface effect
Fig.~\ref{fig:surf}b 
shows results for 16 Cyg A, based on the {\it Kepler} observations
presented by \citet{Metcal2012}.
I made a fit of the observed frequencies to stellar models, using the
\citet{Kjelds2008} surface-term procedure, and subsequently 
determined the best-fitting surface correction according to
Eq.~(\ref{eq:solsurf}).
The dashed and solid lines show the power-law surface term and the
term corresponding to Eq.~(\ref{eq:solsurf}), respectively,
while the symbols show the differences between the stellar and
(uncorrected) model frequencies.
Here the advantage of using the solar fit is less dramatic, but
certainly still significant.
Given that 16 Cyg A is quite similar to the Sun (with $M = 1.1 M_\odot$
and $T_{\rm eff} = 5825\,{\rm K}$) that is probably not surprising.
However, it is interesting that data are now becoming available of
sufficient quality to test more detailed properties of the surface effects
in stars other than the Sun.
Also, the results suggest that there might be an advantage in using 
the solar surface functional form in fits to observed frequencies,
rather than the power law, at least for stars in the vicinity
of the Sun.
In any case a comparison of the results of using these two representations
might provide some measure of the systematic error introduced in the
fit by the use of the surface correction.

It is obvious that the assumption underlying these near-surface corrections,
namely that they resemble the solar correction, is in general questionable,
increasingly so for stars with properties further from those of the Sun.
In fact \citet{Dogan2010} were unable to find a solar-like correction for
the F5 star Procyon A.
Thus some care is required in applying the correction, and further tests
are required to investigate whether it might lead to systematic errors in
the outcome of the fits to the observed frequencies.
It is evident that the broad range of stars observed with {\it Kepler} provides
an excellent basis for such tests.
In the longer run the goal must obviously be to improve the modelling of 
the near-surface layers.
In the solar case some improvement has resulted from the use of
detailed hydrodynamical models of the structure of the outermost layers
\citep[e.g.,][]{Rosent1999}, while the success of using time-dependent
convection modelling in the frequency calculations has so far been somewhat
questionable \citep{Houdek2010, Grigah2012}.

\section{Fitting stellar models}


\subsection{Using scaling relations}


The simplest analysis of asteroseismic data is based on 
the overall properties of the oscillations, typically
the large separation $\Delta \nu$ and the frequency $\nu_{\rm max}$
of maximum power.
Using the scaling relations, Eqs (\ref{eq:delnu}) and (\ref{eq:numax}),
normalizing to solar values $\Delta \nu_\odot$ and $\nu_{\rm max, \odot}$,
we immediately obtain
\begin{eqnarray}
{R \over R_\odot} &=& 
\left( {\nu_{\rm max} \over \nu_{\rm max, \odot} } \right)
\left( {\Delta \nu \over \Delta \nu_\odot } \right)^{-2}
\left( {T_{\rm eff} \over T_{\rm eff, \odot} } \right)^{1/2} \\
{M \over M_\odot} &=& 
\left( {\nu_{\rm max} \over \nu_{\rm max, \odot} } \right)^3
\left( {\Delta \nu \over \Delta \nu_\odot } \right)^{-4}
\left( {T_{\rm eff} \over T_{\rm eff, \odot} } \right)^{3/2} 
\end{eqnarray}
\citep{Kallin2010a, Mosser2010}.
Assuming that $T_{\rm eff}$ has been determined, this provides values of
$R$ and $M$, independently of any models.
However, it should be noticed that $\Delta \nu$ and
$\nu_{\rm max}$ appear to quite high powers,
particularly in the expression for the mass, and hence the results
are sensitive to errors in the observed quantities.

It is obvious that such a determination of
the stellar mass and radius depends on the validity of the assumed scalings.
Since the homologous relation between the relevant stars is not exact,
we may expect departures from the scaling of $\Delta \nu$.
This was investigated by \citet{White2011b}, for stars on the main sequence
and the ascending red-giant branch.
They found a variation of a few per cent which was mainly a function
of effective temperature;
since a significant departure from homology is caused by the varying
depth of the convective envelope, which is largely determined by the
effective temperature, such a dependence is not unexpected.
\citet{Miglio2012} found a difference of about 3 per cent between
$\Delta \nu$ in models of ascending red-giant stars and helium burning
clump stars at given mean density, related to the differences in their
internal structure.
It should also be kept in mind that the near-surface problems discussed in
Section~\ref{sec:nearsurf} have an effect on $\Delta \nu$, given their strong
frequency dependence. 
This may cause systematic errors in the results of the fit.

The scaling of $\nu_{\rm max}$, Eq.~(\ref{eq:numax}), has no
fully secure theoretical basis, since it is not yet possible
to make reliable predictions of the amplitudes of stochastically
excited modes or their dependence on frequency.
However, as mentioned above, the relation has substantial observational
support.
A crucial test of this so-called {\it direct technique} for analysing
the asteroseismic data, or other similar asteroseismic techniques to determine
the stellar radius, is evidently comparison to independent determinations
from the stellar angular diameter and distance.
Such tests appear to confirm the validity of the scaling relations
\citep{Huber2012, Silva2012}.
A similar test of mass determinations for individual stars has 
apparently not been possible.
However, application to red-giant stars in open clusters provides a test
that the resulting masses span the expected small interval
\citep{Basu2011, Miglio2012}.

A more strongly constrained, and hence potentially more precise, determination
of stellar parameters can be obtained by fitting the observed
asteroseismic quantities, together with the effective temperature and
composition, to grids of stellar models.
The determination of $\Delta \nu$ and $\nu_{\rm max}$ for the models
in the grid can be based either on the scaling relations or, in the case
of $\Delta \nu$, on computed frequencies for the models
\citep[e.g.,][]{Stello2009, Basu2010}.
The accuracy of the results is clearly sensitive to the stellar models,
and hence it is important to test the effect on the fits of using
grids computed with different evolution codes
\citep{Gai2011, Basu2012}.
The grid-based fitting was extended to include also the small 
frequency separations by \citet{Quirio2010} in the so-called SEEK 
algorithm, which also included a statistical analysis of the results.

These techniques provide a fast and potentially automated way to obtain
basic stellar properties from the asteroseismic data, and hence they
are well suited to the analysis of large samples of stars, in what has
been called {\it ensemble asteroseismology}%
\footnote{although, according to Douglas Gough, 
{\it sinasteroseismology} might be a more appropriate term}.
This has been extremely valuable for stellar population studies
for main-sequence stars \citep[e.g.,][]{Chapli2011, Verner2011b} and
red giants \citep{Kallin2010b, Mosser2010, 
Hekker2011a, Hekker2011b, Miglio2012}.

\subsection{Fits to individual frequencies}

The basic asteroseismic quantities $\Delta \nu$ and $\nu_{\rm max}$ 
can be determined from data with even rather poor signal-to-noise levels
and, as discussed above, may provide a determination of stellar 
global properties.
However, with the long timeseries obtained from CoRoT and {\it Kepler} it is 
often possible to determine individual frequencies $\nu_{nl}$ 
\citep[e.g.,][]{Appour2008, Appour2012, Barban2009,Campan2011,
Mathur2011}.
This obviously provides far more information about the stellar properties.

The analysis is typically carried out by fitting the frequencies to a set 
of models, characterized by parameters $\{\CP_k\}$ which typically include
the mass, age and initial chemical composition of the star as well as, possibly,
the mixing length and a parameter characterizing convective-core overshoot.
In its simplest form the analysis aims at minimizing 
\begin{equation}
\chi_\nu^2 = {1 \over N - 1}
\sum_{nl} \left( { \nu_{nl}^{\rm (obs)} - \nu_{nl}^{\rm (mod)} 
\over \sigma(\nu_{nl}) }\right)^2 \; ,
\label{eq:chinu}
\end{equation}
as a function of the parameters $\{\CP_k\}$.
Here $N$ is the number of observed frequencies $\nu_{nl}^{\rm (obs)}$,
$\nu_{nl}^{\rm (mod)}$ are the model frequencies,
possibly corrected for near-surface effects according to 
Eq.~(\ref{eq:surface}) and $\sigma(\nu_{nl})$ are the standard errors in the
observed frequencies.
To this may be added a similar term from other observed properties of the star,
such as effective temperature, surface gravity, [Fe/H], and, when available,
radius and luminosity, to obtain the $\chi^2$ which is minimized.

As is common in multi-parameter fits, $\chi^2$ often shows multiple 
local minima,
and a blind search risks landing in what is not the globally smallest value.
Genetic algorithms provide a relatively efficient technique for localizing
the global minimum \citep[e.g.,][]{Metcal2003}.
An implementation for the analysis of solar-like oscillations was presented
by \citet{Metcal2009}; 
this also included a local analysis based on singular-value decomposition (SVD)
to locate more precisely the true minimum, based on the typically relatively
coarsely sampled genetic fit, and determine the statistical properties of
the fit \citep[see also][]{Brown1994}.
This has been implemented in the AMP package%
\footnote{Asteroseismic Modeling Portal; see {\tt http://amp.ucar.edu/}}
for analysis of solar-like oscillation frequencies, which has been made 
available to the asteroseismic community \citep{Woitas2010}.
This is based on the Aarhus stellar evolution code
\citep[ASTEC;][]{Christ2008a}
and pulsation code \citep[ADIPLS;][]{Christ2008b}.


Although the genetic algorithm is effective in localizing the absolute minimum
in $\chi^2$, there may be parameters which provide a fit that differs 
insignificantly from this best fit.
In some cases this may correspond to parameters rather different from those of
the best fit \citep[e.g.,][]{Metcal2010}.
Such non-uniqueness, which is a clear indication of the intrinsic nonlinearity 
of the dependence of the frequencies on the parameters,
requires further study.
It is likely that it may be broken by increased precision in the observations,
an increased range of the observed frequencies or possibly the inclusion
of additional observables in the fit.


An unconstrained fit may yield optimal fits that can be judged to unphysical 
from independent information.
An example is the fit by \citet{Mathur2012}, using AMP, 
to 22 {\it Kepler} stars;
here the initial helium abundance $Y_0$ resulting from the fit in several
cases was substantially below the cosmologically determined primordial
helium abundance $Y_{\rm P} = 0.248$ \citep{Steigm2007}. 
In some cases a value was obtained at the lower limit of the range in
$Y_0$ allowed in the fit.
Although exceptionally interesting, if correct, 
the low value of $Y_0$ is more likely caused by
problems in the fitting or the underlying models.
A related, if less evident, example could be an
inferred $Y_0$ and initial heavy-element abundance $Z_0$ which depart strongly
from the normally assumed relation from Galactic chemical evolution.
Also, for an otherwise sun-like star a value of the inferred mixing length 
very far from the solar calibration might be suspect.

Potential problems of this nature can be avoided by including
prior information in the fit,
although one must obviously be careful not to let the result be too strongly
influenced by perhaps unwarranted prejudice.
Such prior information can be incorporated through a Bayesian approach,
as was also done in the SEEK algorithm \citep{Quirio2010}.
An interesting example, applied to fits of individual frequencies
to grids of stellar models,
was provided by \citet{Gruber2012}.
They demonstrated that this procedure can also be used to suppress the
effect of the surface problems.
A statistical characterization of inferred stellar parameters, using
also Bayesian priors, was obtained by \citet{Bazot2012} 
using a Markov Chain Monte Carlo algorithm in a fit to observations of
$\alpha$ Cen A, although at very considerable computational expense.



In evolved stars the observation of mixed modes provides much more stringent
constraints on the stellar parameters.
The buoyancy frequency (cf.\ Eq.~\ref{eq:buoy}) changes rapidly with age and
so, therefore, do the frequencies of modes with a strong g-mode component.
This provides a very precise determination of the stellar age
\citep[e.g.,][]{Metcal2010}.%
\footnote{It should perhaps be emphasized that the {\it accuracy}
of such determinations still depends sensitively on uncertainties in the
stellar modelling.}
A powerful technique for the analysis of data with mixed modes was presented
by \citet{Deheuv2011} and applied to CoRoT data
\citep[for an application to {\it Kepler} data, see][]{Deheuv2012}.
Also, \citet{Benoma2012} showed how additional information could be obtained
from the detailed properties of the frequencies of such mixed modes.


A major breakthrough in the seismic analysis of red giants has been 
the detection of mixed dipolar modes with a strong g-mode component
\citep{Beck2011, Beddin2011, Mosser2011}.
These appear as a `dipolar forest', a group of peaks around the position
of the dipolar acoustic resonances where the modes have their strongest p-mode
character.
The period spacings between the modes in such groups
reflect the uniform period spacing for pure g modes (cf. Eq.~\ref{eq:gasymp}),
although with departures caused by the interaction with the acoustic cavity.
It is possible to extrapolate the observed period spacings to determine 
the true g-mode spacing $\Delta \Pi_1$ and hence constrain the properties of
the buoyancy frequency in the core of the star \citep[e.g.,][]{Stello2012}.
It was found by \citet{Beddin2011} that $\Delta \Pi_1$ provides a clear 
distinction between stars on the ascending red-giant branch, where the energy
comes solely from the hydrogen-burning shell around an inert helium core,
and the `clump stars', where in addition there is helium fusion near the centre
\citep[see also][]{Mosser2011}.
This distinction comes in part from the fact that the helium-burning stars
have a small convective core, where consequently the buoyancy frequency is
essentially zero \citep{Christ2012a}.
An asymptotic expression, characterizing the frequencies in terms of the
period spacing and the coupling between the gravity-wave and acoustic regions,
was derived by Goupil (in preparation), based on a general asymptotic analysis
\citep{Shibah1979, Unno1979} \citep[see also][]{Christ2012b}.
This was used by \citet{Mosser2012a} for an ensemble analysis of red giants
observed by {\it Kepler}.
Procedures for detailed fits to the observed frequencies of mixed dipolar modes
are under development, and will probably benefit from an application of
this asymptotic expression.

\subsection{Analysis of acoustic glitches}
\label{sec:glitch}

In the analysis of stellar oscillation frequencies information about 
specific features of the star may be obtained through suitable
combinations of frequencies, often based on asymptotic analyses of the
properties of the oscillations.
An important example is the effect of sharp features, i.e., properties in
the star that vary on a scale substantially smaller than the local
wavelength of the oscillations
\citep{Gough1988, Voront1988, Gough1990}.
This causes an oscillatory variation in the frequency,
depending on the phase of the eigenfunction at the location of the feature;
the period of the variation depends on the location of the feature,
while its amplitude, as a function of frequency,
depends on the detailed properties of the feature.
In the case of solar-like oscillations these effects are related to 
{\it acoustic glitches}, with rapid variations in sound speed.
An important example 
arises in the second helium ionization zone, as a result of the
variation in the adiabatic exponent
$\Gamma_1 = (\partial \ln p / \partial \ln \rho)_{\rm ad}$
(the derivative being at constant specific entropy),
and hence in the sound speed, given by $c^2 = \Gamma_1 p/\rho$.
Glitches also occur
at boundaries of convective regions where the temperature gradient
varies rapidly, the sound-speed variation possibly enhanced by variations in
composition.

\begin{figure}
\includegraphics[width=80mm]{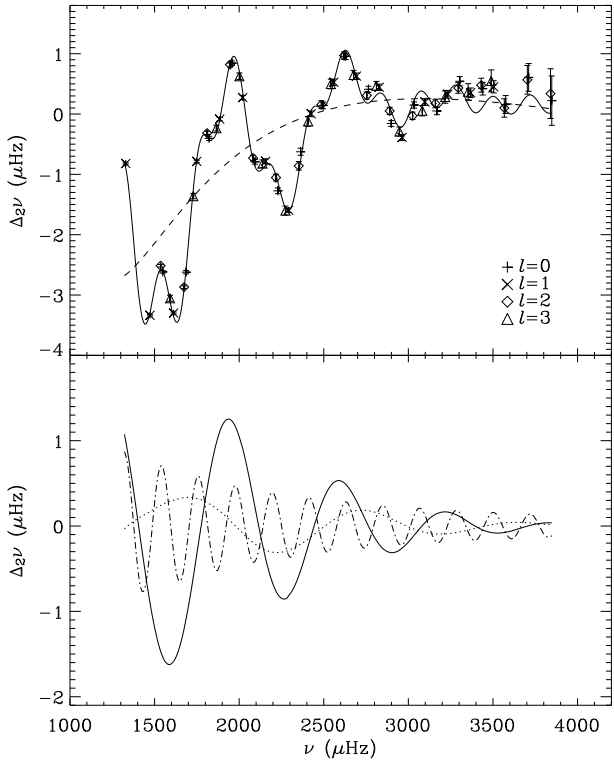}
\caption{
Effects of acoustic glitches on observed solar frequencies.
In the top panel the symbols show second differences obtained from BiSON
observations \citep[see][]{Chapli2002}. 
The solid curve shows the fit consisting of the effect of the glitches from
the ionization zones and the base of the convective envelope,
as well as a slowly varying near-surface contribution
(shown by the dashed line).
The lower panel shows the individual glitch contributions:
the dotted and solid lines show the contributions from the
first and second helium ionization zones, respectively, and the dot-dashed line
shows the contribution from the base of the convective envelope.
From \citet{Houdek2007}.}
\label{fig:sunglitch}
\end{figure}


To isolate the effects of the acoustic glitches slower variations in the
frequencies must be eliminated.
An effective way of doing that is to consider the second differences, as 
a function of mode order:
\begin{equation}
\Delta_2 \nu_{nl} = \nu_{n-1\,l} - 2 \nu_{nl} + \nu_{n+1\,l} 
\label{eq:secdif}
\end{equation}
\citep[e.g.,][]{Gough1990}
or possibly higher differences;
obviously, it is important also to consider the error properties of the
resulting diagnostics.
\citet{Houdek2007} made a careful analysis of the second difference of
solar low-degree frequencies, taking into account the hydrogen and helium
ionization zones and the base of the convective envelope;
their fit, clearly separating these contributions, is shown in 
Fig.~\ref{fig:sunglitch}.
Alternatively, one can consider suitable filtering of the observations
\citep{Perez1994a}
or subtraction of a fitted smooth function from the frequencies
\citep{Montei1994, Montei2005}.

The effects of the second helium ionization obviously depend on the
helium abundance in the stellar convection zone.
This has been used to determine the solar envelope helium abundance from
helioseismic data \citep[e.g.,][]{Voront1991, Antia1994, Perez1994b}.
It was shown by \citet{Perez1998} and \citet{Basu2004} that a similar analysis
is in principle possible for other main-sequence stars, the effect 
increasing with increasing stellar mass.
The properties of the base of the solar convective envelope have
been analysed in considerable detail using the resulting acoustic
glitch, with emphasis on determining the nature of possible convective
overshoot
\citep[e.g.,][]{Basu1994, Montei1994, Roxbur1994, Christ2011}.
A similar analysis is, at least in principle, possible on the basis of
stellar observations of low-degree modes \citep{Montei2000, Mazumd2001}.
\citet{Ballot2004} made a detailed analysis of the potential of determining
the depth of stellar convection zones from the frequency variations
caused by the resulting glitch.
Also, \citet{Mazumd2005} analysed the diagnostic potential of 
the signatures of acoustic glitches.

The effects on the frequencies of acoustic glitches are quite subtle and hence
it is only with the extensive data from CoRoT and {\it Kepler} that it has
been possible to identify them for solar-like oscillations in stars other than
the Sun.
Based on analysis of CoRoT observations by \citet{Carrie2010},
\citet{Miglio2010} found the signature of the glitch caused by the
second helium ionization zone in a red giant.
They used the inferred depth of the ionization zone, together with other
asteroseismic parameters, to obtain a purely seismic determination of the
mass and radius of the star;
however, the data were not yet of sufficient quality to allow a determination
of the envelope helium abundance.
\citet{Mazumd2012a} analysed CoRoT observations of a main-sequence star 
slightly more massive than the Sun and identified a very clear signature
of the second helium ionization zone and a weak and somewhat ambiguous 
signature of the base of the convective envelope.
Very encouraging results were obtained by \citet{Mazumd2012b} for several
stars observed with {\it Kepler}.

There may be circumstances where the effects of acoustic glitches 
might hide or significantly affect other asteroseismic diagnostics.
Thus \citet{Houdek2011} `corrected' solar data for the effect
of the glitches associated with the second helium ionization zone and
the base of the convective envelope before using the resulting glitch-free
frequencies to infer the seismic age and heavy-element abundance of the Sun,
from an asymptotic analysis of the effect of the stellar core.
Of course, the by-products obtained in such a glitch analysis are also
of substantial diagnostic value!

Sharp features in the sound speed and other seismically relevant
quantities are produced by the composition structure resulting from
the evolution of stars with convective cores.
This is particularly dramatic in moderate-mass stars where the
mass contained in the convective core grows with age, resulting
in a composition (and hence sound-speed) discontinuity if diffusion
is neglected.
\citet{Cunha2007} derived a diagnostic applicable to this case and
showed that it provided a measure of stellar age.
This and similar diagnostic quantities were further analysed by
\citet{Cunha2011}.
A similar analysis was carried out by \citet{Silva2011}, demonstrating
the potential of such diagnostics to constrain the size of the convective
core, including the effects of semiconvection which remain a 
serious uncertainty in the modelling of stellar evolution.

With the continuing operations of both CoRoT and {\it Kepler} and the resulting
longer timeseries there is clearly a substantial potential for this type
of analysis.
The depth of convective envelopes is an important parameter in the modelling
of stellar dynamos assumed responsible for stellar magnetic cycles,
while independent determinations of the helium abundance for a range of
stars would obviously be very valuable for studies of the Galactic
chemical evolution and as a constraint on Big-Bang nucleosynthesis.

\section{Inversions}
\label{sec:inv}


The analysis in the preceding section generally characterizes the stellar 
properties in terms of a limited set of parameters.
In particular, in fits such as the minimization of $\chi^2$
(cf. Eq.~\ref{eq:chinu})
the model is characterized by the parameters $\{\CP_k\}$.
Such fits implicitly assume that the physics of the modelling is correct,
which is obviously doubtful.
Indeed, perhaps the most interesting aspect of the asteroseismic analysis 
is to falsify this assumption and, ideally, determine how the modelling
should be improved.

In the simple minimization of $\chi_\nu^2$ an inconsistency in the modelling
is implied if the minimum of $\chi_\nu^2$ is substantially bigger than unity.
However, this is clearly only significant if in fact the errors in the 
observations have been estimated correctly, which in itself is not a trivial
task. 
A single mode with an erroneously low estimated error (or a single
misidentified frequency) could lead to an unrealistically large value
of $\chi_\nu^2$.
Thus care, and perhaps some conservatism, is required in the interpretation
of the results of the fits.

If indeed there is convincing evidence that the large value of $\chi^2$ 
is significant, the task is obviously to determine the cause of the 
discrepancy, in terms of the stellar modelling.
This would benefit greatly from the ability to locate the cause of the
frequency differences between the star and the best model.
Independently of these issues of model fitting, it is clearly of interest
to obtain information about the stellar structure that is not
tied to specific models, as exemplified by the glitch analysis
discussed in Section~\ref{sec:glitch}.
Here I provide a brief discussion of {\it inverse analyses} with this aim.

Such techniques have been extremely successful in helioseismic analysis
of solar oscillation data
\citep[see, for example][for reviews]{Gough1996, Christ2002}.
Here the mode set is sufficiently rich that it is possible to infer 
the sound speed and density in a very large fraction of the solar interior;
although this is typically based on determining corrections to a reference
model, 
the results are generally insensitive to the precise choice of model
\citep{Basu2000}.
The analysis is based on the assumption that the model is sufficiently
close the true solar structure that the frequency differences 
$\delta \nu_{nl}$ between the Sun and the model can be obtained
from an expression linearized in the differences in structure:%
\footnote{Formally, an expression of this form follows from the
fact that adiabatic oscillations satisfy a variational principle.}
\begin{eqnarray}
{\delta \nu_{nl} \over \nu_{nl}} &=&
\int_0^R \left(K_{c^2, \rho}^{nl} {\delta_r c^2 \over c^2} +
K_{\rho, c^2}^{nl} {\delta_r \rho \over \rho} \right) {\dd r} \nonumber \\
&& + Q_{nl}^{-1} \CG(\nu_{nl}) \; .
\label{eq:freqdif}
\end{eqnarray}
Here $\delta_r c^2$ and $\delta_r \rho$ are the differences, at fixed $r$,
between the Sun and the reference model in squared sound speed and density,
and $\CG$ is a function that accounts for the near-surface problems of the
model, with a scale factor $Q_{nl}^{-1}$ that depends on the
inertia of the modes.
The kernels $K_{c^2, \rho}^{nl}$ and $K_{\rho, c^2}^{nl}$ are computed
from the eigenfunctions of the modes of the reference model.

The inverse analyses, based on techniques originally developed in geophysics
\citep[e.g.,][]{Backus1968},
consist of combining relations of the form given in Eq.~(\ref{eq:freqdif})
in such a way as to obtain an estimate of, for example, 
$\delta_rc^2 / c^2$ that is localized near point, $r = r_0$, in the Sun.
If this can be done throughout the star an estimate is obtained of
the sound-speed difference between the Sun and the model.
Given the extremely accurate solar oscillation frequencies spanning
degrees from 0 to at least 200 this can in fact be done with high
precision between $0.07 R$ and $0.95 R$ \citep[e.g.,][]{Basu1997}.
The inferences are obtained as linear combinations of the frequency
differences and can be expressed as averages of the true
sound-speed difference, weighted by averaging kernels
$\CK_{c^2, \rho}(r_0,r)$.
The success of the inversion is characterized by the properties of $\CK$,
particularly its width, and the error in the inferred differences.
The properties of the inversion are typically determined by parameters 
that, for example, control the balance between localizing $\CK$ and 
minimizing the errors.
The choice of these {\it trade-off parameters}, and other aspects of the 
helioseismic inversions, was discussed by \citet{Rabell1999}.

In the stellar case we are typically, for the foreseeable future, restricted
to modes of degree $l = 0{-}3$ and hence the potential for inversion
is much more restricted.
Brief reviews of the related issues were provided by 
\citet{Thomps2002} and \citet{Basu2003},
concentrating on solar-like stars.
Given the limited information it is advantageous to choose a description
of the stellar interior that, as far as possible, depends on a single function 
of position in the star, while still recognizing that the modes observed
are predominantly of acoustic nature.
Using that $c^2 = \Gamma_1 p/\rho$ 
a reasonable choice is the squared isothermal sound speed $u = p/\rho$.
In fact, in much of solar-like stars $\Gamma_1 \simeq 5/3$ and hence
the acoustic properties are fully characterized by $u$.
Assuming furthermore that the equation of state of the stellar matter is
known with sufficient accuracy, $\Gamma_1$ can be found as a function
$\Gamma_1(p, \rho, \{X_i\})$, where $\{X_i\}$ is the composition.
Characterizing the dependence on composition simply by the helium abundance
$Y$ by mass
\footnote{In most cases the heavy-element abundance has a negligible 
effect on the thermodynamic state.}
we can rewrite Eq.~(\ref{eq:freqdif}) as
\begin{eqnarray}
{\delta \nu_{nl} \over \nu_{nl}} &=&
\int_0^R \left(K_{u, Y}^{nl} {\delta_r u \over u} +
K_{Y, u}^{nl} \delta_r Y \right) {\dd r} \nonumber \\
&& + Q_{nl}^{-1} \CG(\nu_{nl}) \; ,
\label{eq:freqdifu}
\end{eqnarray}
where in addition we used the equations of hydrostatic equilibrium and mass
to express the dependence on $\delta_r p$ and $\delta_r \rho$ in terms of
$\delta_r u$.

A second important difference between helioseismic and asteroseismic inverse
analysis is that for stars, unlike the case for the Sun, we do not have 
accurate independent determinations of the mass and radius.
To separate the inverse analysis from the determination of $M$ and $R$
we can rewrite Eq.~(\ref{eq:freqdifu}) in terms of $x = r/R$,
the dimensionless function $\hat u = (R/G M) u$ and the dimensionless 
frequencies $\sigma_{nl} = 2 \pi (R^3/G M)^{1/2} \nu_{nl}$:
\begin{eqnarray}
{\delta \sigma_{nl} \over \sigma_{nl}} &=&
\int_0^1 \left(K_{u, Y}^{nl} {\delta_x \hat u \over \hat u} +
K_{Y, u}^{nl} \delta_x Y \right) {\dd x} \nonumber \\
&& + Q_{nl}^{-1} \hat \CG(\sigma_{nl}) \; .
\label{eq:freqdifus}
\end{eqnarray}
Here I somewhat crudely assumed that the differences in $R$ and $M$ between
the star and the reference model are absorbed in $\hat \CG$.

A final difference between the helioseismic and asteroseismic case concerns
the linear approximation underlying Eq.~(\ref{eq:freqdif}).
It is perhaps reasonable, not least from the tight constraints on mass,
radius, age and luminosity,
to assume that solar models are so close to
the solar structure that the linearized representation of the differences
is sufficiently accurate.
This is not as obviously the case for other stars.
In fact, analyses by Christensen-Dalsgaard \& Thompson (in preparation)
indicate substantial departures from linearity for frequencies of
models differing in mass only by a few percent from the reference model. 
These issues need further investigation before reliable inversions using 
this technique can be carried out.

There is relatively limited experience with this type of inversion as applied
to solar-like stars; 
in particular it is only with the recent space-based data that sufficiently
extensive data are available to make it realistic to consider 
such analyses.
Early tests of inversion of artificial data for low-degree modes were
made by \citet{Gough1993a, Gough1993b}.
\citet{Basu2002} tested inversions for model frequencies 
for reasonably realistic sets of stellar artificial data,
with both models having solar mass and radius.
Based on Eq.~(\ref{eq:freqdifus})
they found that relatively localized averaging kernels could be constructed
in the stellar core, and that the inferred $\delta_r u$ showed clear 
evidence for the core mixing imposed in the test model.
An interesting variant of the linearized inversion, used to infer the
stellar mean density, was discussed by \citet{Reese2012} who applied it to
several sets of observed data.

A conceptually very different technique for asteroseismic determination
of stellar structure has been developed by Roxburgh \& Vorontsov
\citep{Roxbur2002, Roxbur2003b} \citep[see also][]{Roxbur2010}.
This is based on the fact that acoustic modes in the outer parts of the star
can be characterized by a phase function which is independent of the
degree of the mode.
By imposing this constraint, at the observed frequencies, it is possible to
determine the structure of the bulk of the star as parameterized in
terms of a suitable variable, such as the density,
with fitting parameters that are determined iteratively;
it is assumed that in the relevant part of the star, $\Gamma_1 \simeq 5/3$.
Thus the technique does not depend directly on a reference model,
although an initial model is used as a starting point for the iteration.
Applications to artificial data have yielded results remarkably close to
the original model on which the data were based.

\section{Concluding remarks}


The advances in stellar inferences over the last few years, from the
observations made by the CoRoT and {\it Kepler} missions have been remarkable,
far beyond the expectations before the launch of these missions.
In addition to the quality of the data, these advances have been based on
the rapid development of techniques for the analysis of the data. 
It is probably fair to say that the asteroseismic community was not 
entirely prepared for the overwhelming quality, and quantity, of the data
that we have obtained.

While striking results have been obtained, we are not yet quite at the 
point of challenging the modelling of stellar structure and evolution,
with the exception of the evolution of stellar rotation where the theoretical
basis is still highly uncertain.
Such tests and, one might hope, improved understanding of stellar internal
physics are of course a key goal of asteroseismology.
To reach this point the improvements in the data quality and quantity 
resulting from the extension of the missions will certainly be very helpful.
Equally important, however, will be the continued development of the techniques
used to analyse and interpret the observations.
This includes a better statistical analysis of the results, which will be
required to test whether existing models deviate in a significant manner
from the observations.
A key tool in this development will be the application of the techniques
to artificial data;
such tests were carried out before the launch of the missions
\citep[e.g.,][]{Montei2002, Mazumd2005}
but have since largely been ignored in the flush of real data.
Similarly, further work is needed to understand the differences between
results of different stellar evolution and pulsation codes,
as started in the ESTA project in preparation for the CoRoT mission
\citep{Montei2008}.
Finally, it is probably time to consider the application of the inverse
techniques discussed in Section~\ref{sec:inv},
or at least to continue developing and testing these techniques on
the basis of the understanding of the realistic data properties that has
resulted from the observations.
Also, the potential for inversion localized in the stellar core
is undoubtedly improved for stars showing mixed modes, as was demonstrated
by \citet{Lochar2004} in the corresponding case of inversion for rotation.
This deserves further detailed study.

Beyond the present missions there are proposals for additional missions 
combining asteroseismology and the study of extra-solar planets, such as the
Transiting Exoplanet Survey Satellite (TESS) under consideration by NASA
and the PLAnetary Transits and Oscillations of stars 
mission \citep[PLATO;][]{Catala2009} proposed to ESA.
If selected, they will extend the asteroseismic investigations to more
nearby stars in very large parts of the sky, although the duration of the
observations will be shorter than in the case of longest timeseries 
obtained with {\it Kepler}.
Also, the Stellar Observations Network group (SONG) network
\citep[e.g.][]{Grunda2011} will yield radial-velocity observations of
stellar oscillations of even higher asteroseismic quality than the 
{\it Kepler} data, although for a much smaller sample of stars.
Thus it is crucial to select the astrophysically most interesting
targets for SONG,
ideally based on the understanding of stellar properties obtained
from the space missions.
Here, also, the analysis of realistic artificial data is essential.


\acknowledgements
I am very grateful to Markus Roth, and the European Science Foundation,
for the organization of an excellent workshop in beautiful surroundings.
Funding for the Stellar Astrophysics Centre is provided by The Danish 
National Research Foundation. The research is supported by the ASTERISK 
project (ASTERoseismic Investigations with SONG and Kepler) funded by 
the European Research Council (Grant agreement no.: 267864).

\newpage

\end{document}